\documentclass[aps,prd,twocolumn,floatfix, preprintnumbers]{revtex4}
\usepackage[pdftex]{graphicx}
\usepackage[mathscr]{eucal}
\usepackage[pdftex]{hyperref}
\usepackage{amsmath,amssymb,bm,yfonts}
\usepackage[utf8]{inputenc}
\usepackage{color}
\usepackage{ulem}

\newif\ifarxiv
\arxivfalse

\begin		{document}

\title
    {
    Numerical evolution of the interior geometry of charged black holes
    }

\author{Paul~M.~Chesler}
\affiliation
    {Black Hole Initiative, Harvard University, Cambridge, MA 02138, USA}
\email{pchesler@g.harvard.edu}

\date{\today}

\begin{abstract}
        Previously, we developed a late time approximation scheme to study the interior geometry of black holes.  In the present paper 
        we test this scheme with numerical relativity simulations.  In particular, we present numerical 
        relativity simulations of the interior geometry of charged spherically symmetric two-sided black holes with a spacelike singularity at $r = 0$.
        Our numerics are in excellent agreement with the late time approximation.  
         We also demonstrate that the geometry near $r = 0$ is a scalarized Kasner geometry and compute the associated Kasner exponents.
\end{abstract}

\pacs{}

\maketitle
\parskip	2pt plus 1pt minus 1pt

\section{Introduction}

An interesting question in General Relativity is what is the final state of gravitational collapse?  When a black hole
forms, the external geometry relaxes to the Kerr-Newman solution.  However, the interior geometry is not unique and depends on initial conditions. What then are the universal features of final state interior geometry? A natural guess is the existence and structure of singularities.

In asymptotically flat space one universal interior feature is a null singularity at the Cauchy horizon (CH), located at advanced time $v = \infty$ \footnote{This need not happen in de Sitter space \cite{Cardoso:2017soq}.  However, quantum effects may still lead to a singular 
CH  \cite{Hollands:2019whz}.}.  There are essentially two ingredients required to reach this conclusion.  
The first is Price's Law \cite{Price:1971fb,Price:1972pw}.  A generic localized perturbation of the external geometry results in 
a wave packet of outgoing radiation propagating to $r = \infty$.  This wave packet will continuously scatter off the black hole's gravitational potential, resulting in a small influx of
radiation into the horizon.  Price reasoned the influx decays like $v^{-p}$ where the power $p$ depends on both the spin of the field and its angular momentum.  The second
ingredient is the well-known exponential blueshift near the CH. Observers can reach the CH in a finite proper time, meaning they can observe the entire evolution of the outside universe in a finite proper time.  In their reference frame   
the Price Law influx appears unboundedly blueshifted as the CH is approached.
Numerous studies have confirmed the exponential blueshift of Price Law tails leads to a null singularity at the CH \cite{Penrose:1968ar,Simpson:1973ua,HISCOCK1981110,PhysRevD.20.1260,Poisson:1989zz,PhysRevD.41.1796,PhysRevLett.67.789,0264-9381-10-6-006,Brady:1995ni,Burko:1997zy,Hod:1998gy,Burko:1997fc,10.2307/3597235,Dafermos:2017dbw,Ori:2001pc,Ori1997,Burko:2016uvr,Dias:2018ynt} %

Recently we argued Price Law tails and the exponential blueshift also necessitate the existence of spacelike singularities in one-sided black holes coupled to a scalar field \footnote{See Ref.~\cite{VandeMoortel:2019ike} for recent proof that the CH cannot close off the spacetime in spherically symmetric one-sided black holes}.  This was 
done for both for spherically symmetric charged black holes \cite{Chesler:2019tco} and for neutral rotating black holes without any symmetry \cite{Chesler:2019pss}.  
Both analyses relied on a late time expansion with expansion parameter $e^{-\kappa v}$, with $\kappa$ the surface gravity of the inner horizon.  The existence of this expansion parameter is intimately tied to the exponential blueshift
near the CH.  Both analyses found that the Kretschmann scalar $K \equiv R^{\mu \nu \alpha \beta}R_{\mu \nu \alpha \beta}$ diverges near $r = 0$ like
\begin{equation}
\label{eq:K}
K \sim r^{-2 \alpha v} e^{2 \kappa v},
\end{equation}
where $\alpha>0$ is a constant related to Price Law influxes.  The strength of the singularity at $r = 0$ increases with $v$
due to buildup of a singular cloud of scalar radiation near $r = 0$ sourced by Price's Law.

\begin{figure}[ht]
	\includegraphics[trim= 0 0 0 0 ,clip,scale=0.3]{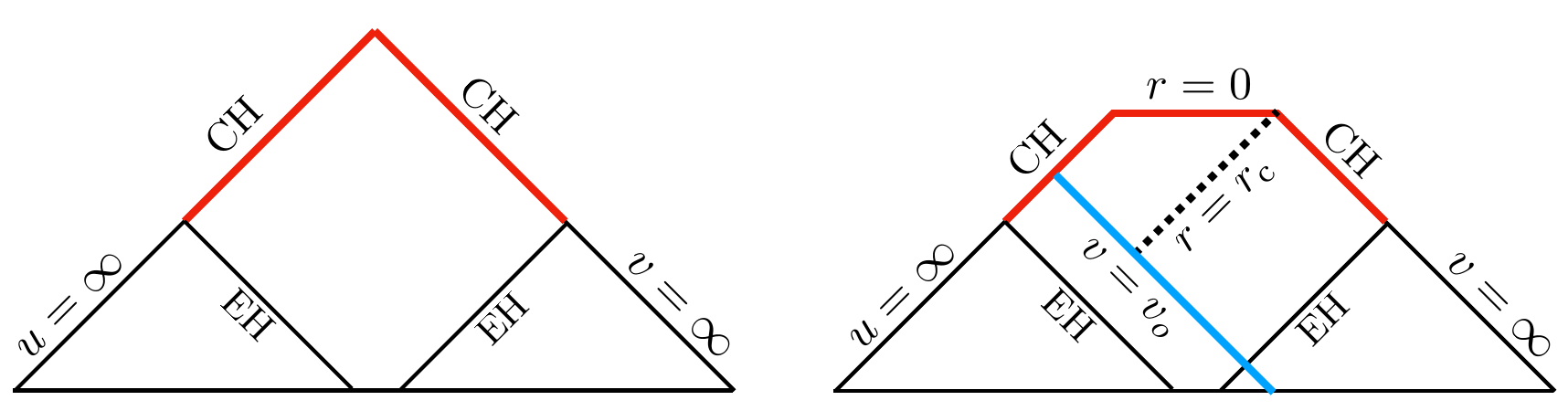}
	\caption{Penrose diagrams showing two possible two-sided black holes. 
		For weakly perturbed initial data (left panel), the geometry only contains null singularities at the CH.  With large perturbations (right panel)
		there can also be a spacelike singularity at $r = 0$.  
	}
	\label{fig:twocases}
\end{figure}

Our primary goal in this paper is to bolster the validity of the late time approximation scheme employed in Ref.~\cite{Chesler:2019tco}
with numerical simulations.  A simple model to study is that of spherically symmetric charged black holes coupled to a real scalar field.
In order to have a non-trivial electromagnetic field strength tensor, charged black holes in this model must be two-sided, meaning the geometry
must contain two separate asymptotically flat regions.  
We note that while Ref.~\cite{Chesler:2019tco} employed charged scalar fields, their analysis can trivially be extended to real scalar fields.
We also emphasize that the analysis in Ref.~\cite{Chesler:2019tco} is local and does not depend on the topology of the black hole.

Fig.~\ref{fig:twocases} shows two possible Penrose diagrams for two-sided black holes (see e.g.  Refs.~\cite{Kommemi:2011wh,Luk:2017jxq}).
The ingoing branch of the CH is located at advanced time $v = \infty$ while the 
outgoing branch is located at retarded time $ u =\infty$.  
Suppose initial data is specified on some Cauchy surface.
For weakly perturbed initial data (i.e. that close to the 
Reissner-Nordstr\"om (RN) solution), the geometry only contains a null singularity on the CH \cite{Dafermos:2012np}.  Such a scenario 
is depicted in the left panel of Fig.~\ref{fig:twocases}.  Ref.~\cite{Dafermos:2012np}
demonstrated the areal radius $r_{\rm CH}$ of the CH  satisfies $r_- -\varepsilon< r_{\rm CH} < r_-$ where 
\begin{equation}
\label{eq:rm}
r_- \equiv M - \sqrt{M^2 - Q^2},
\end{equation}
is the RN inner horizon radius, with $M$ and $Q$ the black hole mass and charge, and $\varepsilon \to 0$ characterizes the size of the initial perturbations of the RN geometry.
However, as the size of initial perturbations is increased, there is no reason to expect $r_{\rm CH} \approx r_-$.  
Indeed, numerical simulations of two-sided black holes with large perturbations indicate the CH contracts to $r = 0$, at which point it meets a spacelike singularity \cite{Brady:1995ni,Burko:1997zy}. This scenario is depicted in the right panel of 
Fig.~\ref{fig:twocases}.  In this case, at large but fixed $v$ the geometry is regular until $r = 0$. 

The late time expansion of Ref.~\cite{Chesler:2019pss} employs a null slicing, 
where initial data is specified on 
some asymptotically late time null surface  $v = v_o$, as depicted in the right panel of Fig.~\ref{fig:twocases}.  
A singular right moving branch of the CH then requires the initial scalar field data to be singular at $r= r_{\rm CH}(v_o)$. 
Within the late time approximation scheme, the future evolution of $r_{\rm CH}$ --- specifically whether it contracts to $r = 0$ or not --- depends on
the initial value of $r_{\rm CH}$.  If $r_{\rm CH} < r_{\rm c}$, then the CH must contract to $r = 0$ in some finite time.
The critical null surface $r_{\rm c}$, which is also shown in the right panel of Fig.~\ref{fig:twocases}, scales like $r_{\rm c} \sim v^{1/2-p}$.  
Unless explicitly stated otherwise, throughout this paper we shall assume the initial data is regular at $r > 0$.

We use infalling Bondi-Sachs coordinates and rederive the late time approximation scheme employed in Ref.~\cite{Chesler:2019tco}.  
Bondi-Sachs coordinates yield a somewhat simpler and more transparent analysis than the coordinate system used in Ref.~\cite{Chesler:2019tco}. 
As found in Ref.~\cite{Chesler:2019tco}, we find that the  geometry contains a null singularity at $v = \infty$ and a spacelike singularity at $r = 0$, with the curvature near $r = 0$ given by (\ref{eq:K}).  Moreover, near the spacelike singularity the geometry 
is that of a scalarized Kasner geometry.  Additionally, \textit{all} time-like curves inside $r = r_-$ end on a singularity within proper time $\Delta \tau \lesssim e^{-\kappa v/2}$.  This time scale merely reflects the exponential blueshift near the CH and indicates that for time-like observers, the classical geometry
effectively ends at $r = r_-$.
We then verify the validity of our late time approximation with numerical simulations.

An outline of the remainder of our paper is as follows.  In Sec.~\ref{eq:einsteinscalarsystem}
we present the system we study.  In Sec.~\ref{sec:latetimeapprox} we 
derive late time solutions to the equations of motion.  In Sec.~\ref{sec:numerics} we present numerical
solutions and compare them to our late time asymptotics.  Finally, we discuss our results in
Sec.~\ref{eq:discussion}.

\section{The Einstein-Maxwell-Scalar system}
\label{eq:einsteinscalarsystem}

We consider the dynamics of 
spherically symmetric charged black holes with a massless real scalar field $\Psi$. 
Einstein's equations, Maxwell's equations,
and the Klein-Gordon equation read,
\begin{subequations}
	\label{eq:eqm}	
	\begin{eqnarray}
	\label{eq:einstein}
	R_{\mu \nu} + {\textstyle \frac{1}{2}} R g_{\mu \nu} &=& 8 \pi (T_{\mu \nu} + \mathcal T_{\mu \nu}), \\ \label{eq:maxwell}
	\nabla_\mu F^{\mu \nu} &=& 0, \\
	\label{eq:scalar}
	\nabla^2 \Psi &=& 0,
	\end{eqnarray}
\end{subequations}
where $\nabla$ is the covariant derivative and 
\begin{subequations}
	\label{eq:stresses}
	\begin{align}
	T_{\mu \nu} &=  \nabla_\mu \Psi \nabla_\nu \Psi - g_{\mu \nu} (\nabla \Psi)^2, 
	\\ \label{eq:emstress}
	\mathcal T_{\mu \nu} &={\textstyle \frac{1}{4 \pi}}\left (-F_{\mu \beta}F^{\beta}_{\ \nu} - {\textstyle \frac{1}{4}} g_{\mu \nu} F_{\alpha \beta}F^{\alpha \beta} \right),
	\end{align}
\end{subequations}
are the scalar and electromagnetic stress 
tensors, respectively.

We employ infalling Bondi-Sachs coordinates where the 
metric takes the form \cite{Madler:2016xju}
\begin{equation}
\label{eq:metric}
ds^2 = e^{2 B}[-2 V dv^2 + 2 dr dv] + r^2  [d \theta^2 + \sin^2 \theta d\phi^2],
\end{equation} 
with $v$ advanced time and $r$ the areal radial coordinate.
Outgoing radial null geodesics satisfy
\begin{equation}
\label{eq:outgoing}
\frac{dr}{dv} = V,
\end{equation}
while infalling radial null geodesics satisfy 
\begin{equation}
\label{eq:infalling}
v = {\rm const.}
\end{equation}
It will be useful below to define directional derivative operators
along both infalling and outgoing null geodesics,
\begin{align}
&'\equiv \partial_r, &&
d_+ \equiv \partial_v + V \partial_r.&
\end{align}

With spherical symmetry and our metric ansatz (\ref{eq:metric}), Maxwell's equations (\ref{eq:maxwell}) are solved by the gauge field $A_{\mu} = \{ \Phi,0,0,0\}$ where the potential $\Phi$ satisfies 
\begin{equation}
\label{eq:potential}
\Phi' = \frac{Q e^{2 B}}{ r^2},
\end{equation} 
with $Q$ the charge of the black hole.  Substituting (\ref{eq:potential}) into (\ref{eq:emstress}) we conclude
\begin{equation}
\label{eq:emstress2}
\mathcal T^\mu_{\ \ \nu} = \frac{Q^2}{8 \pi r^4} {\rm diag}[-1,-1,1,1].
\end{equation}
It follows that dynamics of $A_\mu$ decouple from the Einstein-scalar system. 

With the electromagnetic stress tensor (\ref{eq:emstress2}), Einstein's equations (\ref{eq:einstein})
and the Klein-Gordon equation (\ref{eq:scalar}) reduce to 
\begin{subequations}
	\label{eq:einsteinscalar}
	\begin{eqnarray}
	\label{eq:Beq}
	0 &=& B' - 2 \pi r \Psi'^2, \\
	\label{eq:Veq}
	0&=& (r V)' - {\textstyle \frac{1}{2}} e^{2 B} \left (1 - {\textstyle \frac{Q^2}{r^2}} \right ), \\
	\label{eq:constraint}
	0&=&\partial_v V - 2 V d_+ B + 4 \pi r (d_+ \Psi)^2, \\
	\label{eq:scalar2}
	0&=& (r d_+ \Psi)' + V \Psi'.
	\end{eqnarray}
\end{subequations}
Eqs.~(\ref{eq:einsteinscalar}) have a nested linear structure.  Given $\Psi$ on some $v = {\rm const.}$ surface, the Einstein eqution (\ref{eq:Beq}) can be integrated inwards to find $B$.  With $\Psi$ and $B$ known, the Einstein eqution (\ref{eq:Veq}) can be integrated inwards to find $V$.  With $\Psi$, $B$
and $V$ known, the Klein-Gordon equation (\ref{eq:scalar2}) can be integrated inwards to find $d_+ \Psi$.  With $\Psi$, $V$ and $d_+ \Psi$ known, one can compute $\partial_v \Psi = d_+ \Psi - V \Psi'$ and march forward in time.  To perform this procedure one must specify boundary conditions for $B$, $V$ and $\Psi$.  These boundary conditions are not independent as they must satisfy Eq.~(\ref{eq:constraint}).  Note Eq.~(\ref{eq:constraint}) is a radial constraint equation: if (\ref{eq:constraint}) is satisfied at one value of $r$, then the remaining equations guarantee it is satisfied at all values of $r$.
Hence Eq.~(\ref{eq:constraint}) can be implemented as a boundary condition at some fixed $r$.
A simple choice is to employ Eq.~(\ref{eq:constraint}) to dynamically evolve the value of $V$ on some $r = {\rm const.}$ surface.

Following Ref.~\cite{Chesler:2019pss}, we are interesting in solving the Einstein-scalar system (\ref{eq:einsteinscalar}) at asymptotically late times $v$ (i.e. near the infalling branch of the CH shown in Fig.~{\ref{fig:twocases}). 
Additionally, we restrict our attention to $r  \leq r_{\rm max}$ for some $r_{\rm max} < r_-$.  
Why not simply integrate all the way out to $r = \infty$?  Firstly, numerical simulations indicate the 
geometry at $r > r_-$ simply relaxes to the relaxes to the RN solution (see e.g. \cite{Chesler:2019pss}).  Second,  numerically integrating Einstein's equations across $r= r_-$ is challenging due to shocks which form at $r = r_-$ \cite{Marolf:2011dj,Eilon:2016osg,Chesler:2018hgn,Burko:2019fgt}.  
Nevertheless, at late times one can piece the geometry together with suitable boundary conditions at $r = r_{\rm max}$. In Sec.~\ref{sec:latetimeapprox} we shall find that the solutions at $r < r_{\rm max}$ do not depend on the precise choice of $r_{\rm max}$.

What are the appropriate boundary conditions at $r = r_{\rm max}$ at asymptotically late times?
One boundary condition  simply comes from Price's Law.  At asymptotically late times Price's Law dictates that the scalar field at $r = r_{\rm max}$ decays like,
\begin{equation}
\label{eq:Price}
\partial_v \Psi|_{r = r_{\rm max}} = \frac{A v^{-p}}{r_{\rm max}},
\end{equation}
for some amplitude $A$ and power $p$.  As we shall see below, the factor of $1/r_{\rm max}$ 
accounts for the fact that $\partial_v \Psi \sim 1/r$.
For a spherically symmetric real scalar field $p = 4$ \cite{Price:1971fb,Price:1972pw,Dafermos:2003yw,Donninger:2009tw,metcalfe2011prices,Angelopoulos:2016wcv}.
However, it will be useful to leave $p$ arbitrary. 

\begin{figure}[ht]
	\includegraphics[trim= 0 0 0 0 ,clip,scale=0.275]{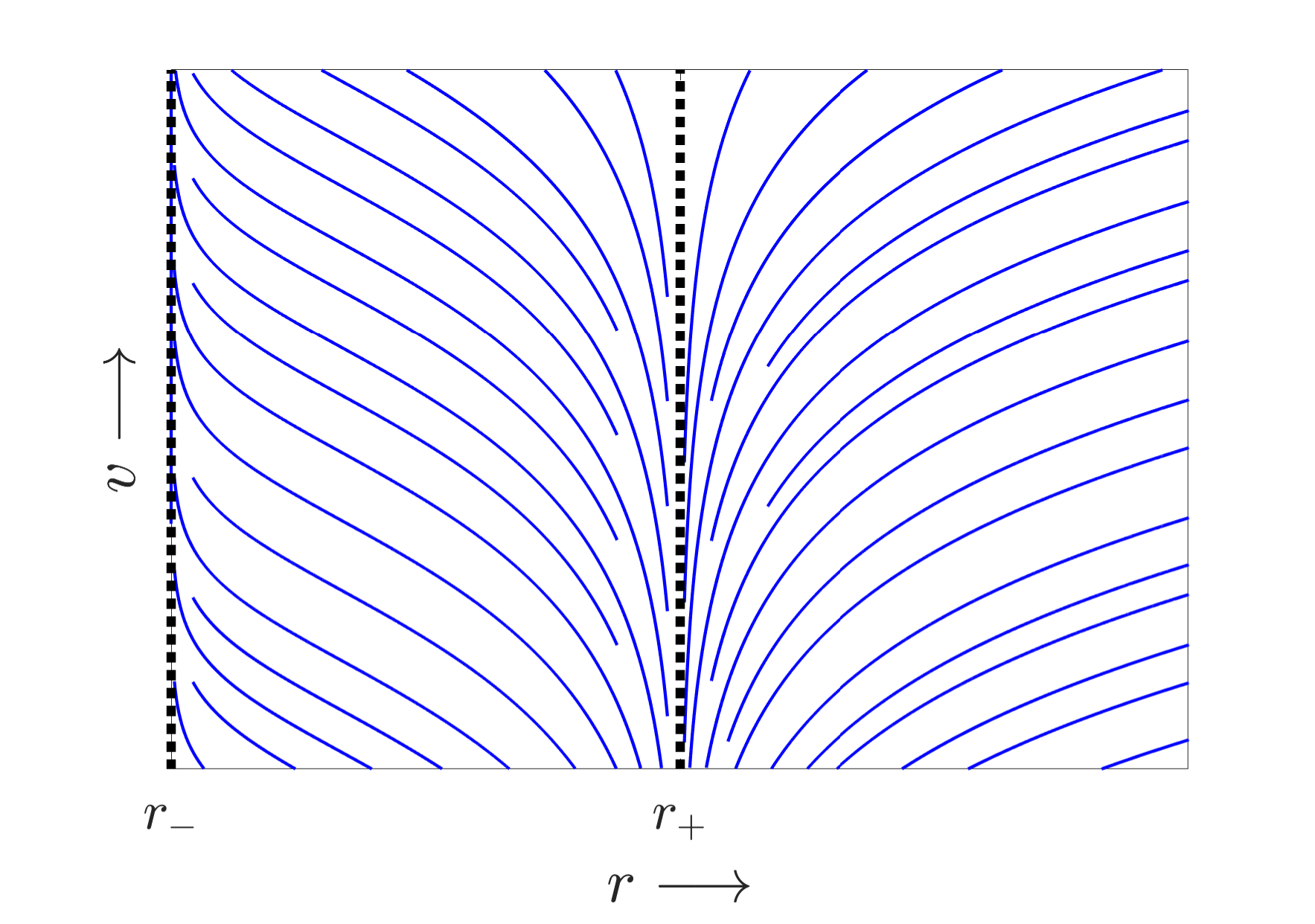}
	\caption{A sketch of a congruence of outgoing null geodesics in the RN geometry.  Outgoing geodesics outside the horizon escape to $r = \infty$ as $v \to \infty$.  
	Outgoing null geodesics at $r_- < r < r_+$ asymptote to $r = r_-$ as $v \to \infty$.
	}
	\label{fig:congruence}
\end{figure}

Our second boundary condition is 
\begin{equation}
\label{eq:blueshift}
\partial_v B|_{r = r_{\rm max}} = -\frac{\kappa }{2},
\end{equation}
where
\begin{equation}
\label{eq:kappa}
\kappa =  \frac{Q^2 - M r_-}{r_-^3},
\end{equation}
is the surface gravity at $r = r_-$ of the associated RN solution (i.e. that with the same mass $M$ and charge $Q$).  

Where does the boundary condition (\ref{eq:blueshift}) come from?
Firstly, assuming the geometry at $r > r_-$ relaxes to the RN solution 
at late times, the metric at $r > r_-$ is given by $B \approx 0$ 
and $2 V \approx 1 - \frac{2 M}{r} + \frac{Q^2}{r^2}$.  The event horizon is located at $r_+ \equiv M + \sqrt{M^2 - Q^2}$.  In Fig.~\ref{fig:congruence} we sketch a congruence of outgoing null geodesics in the RN geometry.  All outgoing null geodesics between $r_-$ and $r_+$
asymptote to $r_-$ as $v \to \infty$.  This means that any outgoing radiation at $r_- < r < r_+$, which must exist due to scattering of Price Law influxes, becomes localized to a ball whose surface approaches $r = r_-$ as $v \to \infty$.  Correspondingly, $\Psi'$ must grow unboundedly large at $r = r_-$ as $v \to \infty$.  Eq.~(\ref{eq:Beq}) implies $B = \int dr 2 \pi r (\Psi')^2$, which means that $B$ must abruptly decrease across $r_-$ with the magnitude of the effective discontinuity growing with $v$.  In other words, an effective shock in $B$ forms at $r = r_-$
\cite{Marolf:2011dj,Eilon:2016osg}.  The Einstein-scalar system can be solved 
analytically near $r = r_-$ using geometric optics \cite{Chesler:2019tco,Chesler:2018hgn}.  Doing so shows that just inside $r = r_-$, $B$ decreases in accord with Eq.~(\ref{eq:blueshift}) 
\footnote{We note the analyses of Refs.~\cite{Chesler:2019tco,Chesler:2018hgn} employ the affine parameter $\lambda$ of infalling null geodesics as a radial coordinate.  This is related to 
$B$ via $\frac{\partial r}{\partial \lambda} = e^{-2B}$}.  

Physically, the boundary condition (\ref{eq:blueshift}) simply encodes the exponential blueshift incurred near the CH.  It implies that clocks belonging to observers attempting to cross the CH run exponentially slows than those of the outside universe.  Moreover, the boundary condition (\ref{eq:blueshift}) 
is necessary for mass inflation \cite{Poisson:1989zz,PhysRevD.41.1796} to occur.
Additionally, note the boundary conditions (\ref{eq:Price}) and (\ref{eq:blueshift}) are strictly 
valid near the CH.  Both boundary conditions presumably receive corrections 
suppressed by inverse powers of $v$.

Boundary data for $V$ must be dynamically determined by integrating the 
radial constraint equation  (\ref{eq:constraint}) at $r = r_{\rm max}$.
Using (\ref{eq:Beq}) and the boundary conditions (\ref{eq:Price}) and (\ref{eq:blueshift}), the radial constraint equation becomes
\begin{equation}
\label{eq:constaint2}
\left [\partial_v V + \kappa V + 4 \pi  \left ( A^2 v^{-2 p}/r + 2 V A v^{-p} \partial_r \Psi \right ) \right ]|_{r = r_{\rm max}} = 0.
\end{equation}
Hence,  $V|_{r = r_{\rm max}}$ satisfies an ODE in $v$.  Because of this, our dynamical evolution variables are the  scalar field $\Psi$ and the boundary value $V|_{r = r_{\rm max}}$,

\section{Late time approximation}
\label{sec:latetimeapprox}

Ref.~\cite{Chesler:2019tco} solved the Einstein-scalar system with a late time expansion, meaning in the limit $v \to \infty$.  In this section we repeat the analysis 
of Ref.~\cite{Chesler:2019tco} verbatim, albeit in Bondi-Sachs coordinates.
Why?  The analysis of Ref.~\cite{Chesler:2019tco} is  simpler and more transparent in
infalling Bondi-Sachs coordinates.  In particular, in Bondi-Sachs coordinates 
the approximation scheme employed in Ref.~\cite{Chesler:2019tco}
merely boils down to neglecting the terms in 
Eq.~(\ref{eq:Veq}) proportional to $e^{2 B}$.  
Why is this justified?
Eq.~(\ref{eq:Beq}) implies $B$ can only decrease as $r$ decreases.  Together with the blueshift boundary conditions (\ref{eq:blueshift}), this means 
\begin{equation}
\label{eq:Bbound}
e^{2B} \lesssim e^{-\kappa v},
\end{equation}
everywhere inside $r \leq r_{\rm max}$.  With this approximation
Eq.~(\ref{eq:Veq}) becomes
\begin{equation}
\label{eq:Veq2}
(r V)' = 0.
\end{equation}
Note that the neglected terms, ${\textstyle \frac{1}{2}} e^{2 B} \left (1 - {\textstyle \frac{Q^2}{r^2}} \right )$, naively become large when $r \sim e^{-\kappa v}$.
However, we shall see below that when $r \lesssim v^{1/2 - p}$, we have $e^{2B} \lesssim r^{\alpha v} e^{\kappa v}$ for some constant $\alpha > 0$.  This means that at late enough times the neglected terms are order $e^{-\kappa v}$ everywhere, including near $r = 0$.

\subsection{Solutions}
\label{sec:sols}

Eq.~(\ref{eq:Veq2}) can be integrated to yield
\begin{equation}
\label{eq:Vsol}
V(v,r) = -\frac{\zeta(v)}{r},
\end{equation}
with constant of integration $\zeta(v)$.  $\zeta$ can be 
determined from the radial constraint equation (\ref{eq:constaint2}).  Assuming $\partial_r \Psi$ remains bounded at $r = r_{\rm max}$, Eqs.~(\ref{eq:constaint2}) and (\ref{eq:Vsol}) imply
\begin{equation}
\label{eq:zeta}
\zeta(v) = \frac{4 \pi A^2}{\kappa} v^{- 2 p}  + \mathcal O(v^{-2p - 1}).
\end{equation}

We note that with $V$ given by Eqs.~(\ref{eq:Vsol}) and (\ref{eq:zeta}), the outgoing null geodesic equation (\ref{eq:outgoing}) is solved by 
\begin{equation}
\label{eq:outgoingsol}
r^2 =\frac{8 \pi  A^2}{\kappa(2 p - 1)}  v^{1 - 2 p} + \rm const.
\end{equation}
Depending on the constant of integration, outgoing geodesics either terminate at the CH, with a finite value of $r$, or plunge into $r = 0$ in a finite time $v$.  The {critical} geodesic,
which only reaches $r = 0$ at $v = \infty$, is given by
\begin{equation}
\label{eq:rc}
r_{\rm c}^2 = \frac{8 \pi A^2}{\kappa(2 p - 1)}  v^{1 - 2 p}.
\end{equation}
This geodesic is shown in the right panel of Fig.~\ref{fig:twocases}.

With $V$ given by Eqs.~(\ref{eq:Vsol}) and (\ref{eq:zeta}),
the Klein-Gordon equation (\ref{eq:scalar2}) is a {decoupled linear} PDE for $\Psi$,
\begin{equation}
\label{eq:scalar3}
(r \partial_v \Psi)' - \frac{\zeta(v)}{r} (r  \Psi')' = 0.
\end{equation}
Defining the ``energy" density $\mathcal E$ and ``energy" flux $\mathcal S$,
\begin{align}
\label{eq:conserved}
&\mathcal E \equiv r (\Psi')^2, &&
\mathcal S \equiv \textstyle \frac{r^2}{\zeta} (d_+ \Psi)^2 - \zeta (\Psi')^2, &
\end{align}
it is easy to see  that Eq.~(\ref{eq:scalar3}) implies the conservation equation,
\begin{equation}
\partial_v \mathcal E + \partial_r \mathcal S = 0.
\end{equation}
Since both the explicit time dependence in the Klein-Gordon equation (\ref{eq:scalar3}) and the Price Law boundary condition (\ref{eq:Price}) 
are arbitrarily slowly varying at late times, it is reasonable to surmise that $\partial_v \mathcal S \to 0$ as $v \to \infty$.
In other words, the flow of energy should approach a steady-state at late times.

We do not know how to compute the general solution to Eq.~(\ref{eq:scalar3}) analytically.  Nevertheless, approximate 
solutions can easily be obtained.
Away from $r = 0$ and at late times we can neglect the last term in Eq.~(\ref{eq:scalar3}), meaning Eq.~(\ref{eq:scalar3})  becomes
\begin{equation}
\label{eq:Psieq1}
(r \partial_v \Psi)' = 0.
\end{equation}
With the Price Law boundary condition (\ref{eq:Price}), the  solution to (\ref{eq:Psieq1}) reads
\begin{equation}
\label{eq:awayfromorigin}
\Psi(v,r) = \frac{A v^{1-p}}{r(1-p)} + f(r),
\end{equation}
where $f(r)$ depends on initial conditions.  
Note $\Psi$ does not 
depend on $r_{\rm max}$, which justifies the factor of $1/r_{\rm max}$ 
in the Price Law boundary condition (\ref{eq:Price}). 
The energy flux associated with (\ref{eq:awayfromorigin}) reads
\begin{equation}
\mathcal S = \frac{\kappa}{4 \pi} + O(v^{-p}),
\end{equation}
which, as anticipated, is approximately constant.
 
\begin{figure*}[ht]
	\includegraphics[trim= 0 0 0 0 ,clip,scale=0.4]{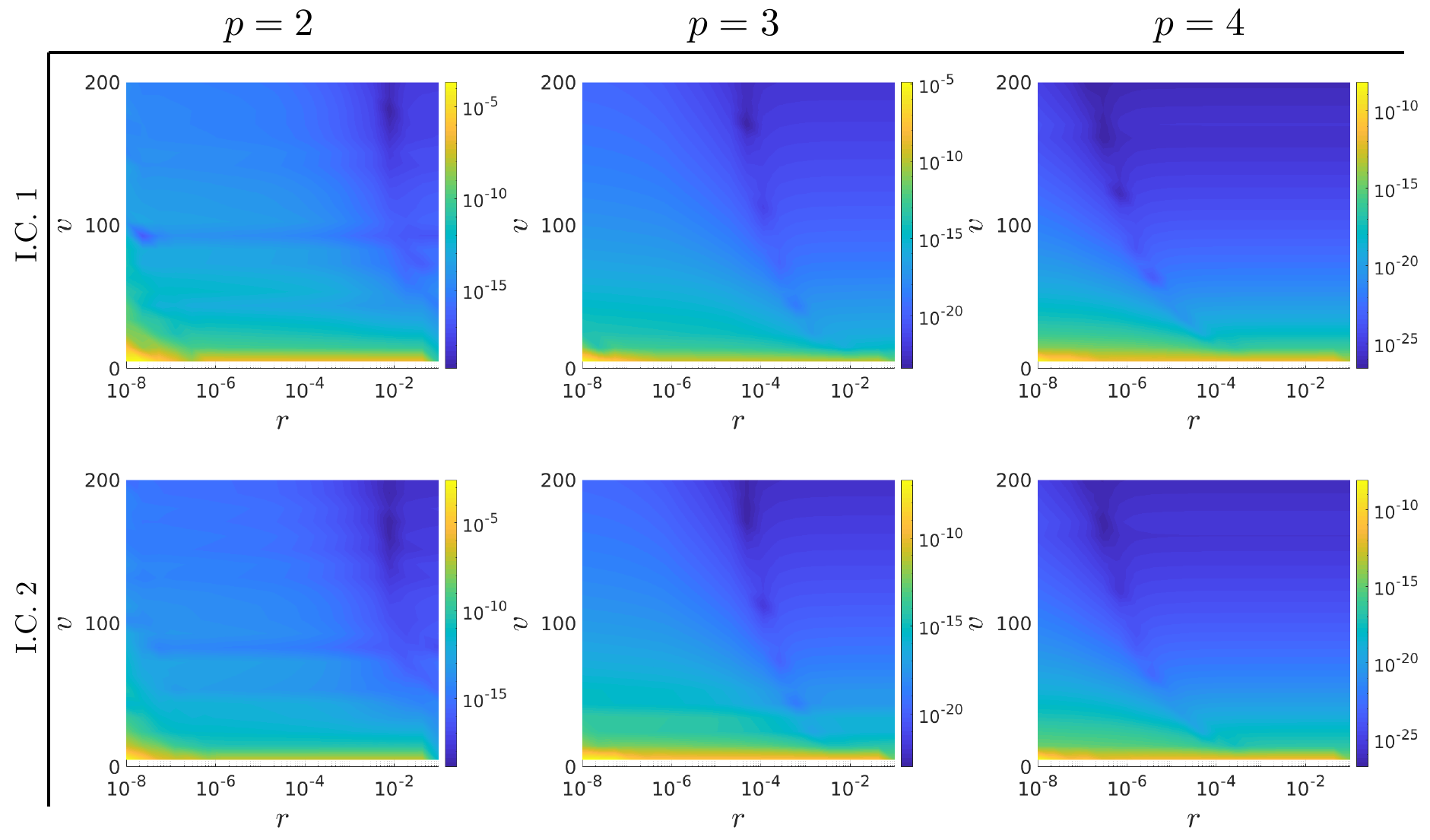}
	\caption{
		Violations $| {\rm Err}|$ of the radial constraint equation (\ref{eq:Erreq})
		for both sets of initial data with $p = 2,3$ and $4$. The fact that $\rm Err \ll 1$ indicates our numerics well-approximate the continuum limit.
	}
	\label{fig:Convergence}
\end{figure*}

Conversely, at sufficiently small $r$ Eq.~(\ref{eq:scalar3}) can be solved with the Frobenius expansion
\begin{equation}
\label{eq:Frobenius}
\Psi(v,r) =  \log r \sum_{n = 0} \Psi_{(n)}(v) {\textstyle \left (\frac{r}{r_{\rm c}} \right)^{2n}} + \sum_{n = 0} \psi_{(n)}(v) {\textstyle \left (\frac{r}{r_{\rm c}} \right)^{2n}.}
\end{equation}
All coefficients $\Psi_{(n)}$ and $\psi_{(n)}$ with $n\geq 2$
are determined by $\Psi_{(0)}$ and $\psi_{(0)}$.  The time dependence of $\Psi_{(0)}$ and $\psi_{(0)}$  is constrained by the quasi steady-state condition $\partial_v \mathcal S = 0$, which near $r = 0$ requires 
\begin{align}
\partial_v [\Psi_{(0)} \partial_v \Psi_{(0)}] = 0, &&
\partial_v [\Psi_{(0)} \partial_v \psi_{(0)}] = 0.&
\end{align}
These equations are solved by 
\begin{align}
\label{eq:Psi0}
&\Psi_{(0)} \sim  \sqrt{v}, &&
\psi_{(0)} \sim  \sqrt{v}.&
\end{align}
Moreover,
all coefficients $\Psi_{(n)}$ and $\psi_{(n)}$ scale like $\sqrt{v}$
as $v \to \infty$.  Presumably,  this means the expansion (\ref{eq:Frobenius}) is well-behaved when $r \lesssim r_{\rm c}$.

Evidently, near $r = 0$ the scalar field 
diverges like
\begin{equation}
\label{eq:nearoriginscalar}
\Psi \sim  \sqrt{v} \log r.
\end{equation}
At what radius does the solution (\ref{eq:awayfromorigin}) match onto the scaling (\ref{eq:nearoriginscalar})? 
The approximation that went into obtaining 
equation of motion (\ref{eq:Psieq1}) 
 breaks down when $r \sim r_{\rm c}$.  
This suggests the scalar field transitions from (\ref{eq:awayfromorigin}) to (\ref{eq:nearoriginscalar}) when $r \sim r_{\rm c}$.
Our numerical simulations presented in Sec.~\ref{sec:numerics} are consistent with this.
It is noteworthy that the $\sqrt{v}$ growth in Eq.~(\ref{eq:nearoriginscalar}) is the same for all $p$.
However, the domain of applicability of Eq.~(\ref{eq:nearoriginscalar}),  $r \lesssim r_{\rm c} \sim v^{1/2 - p}$, is sensitive to the value of $p$.

Finally, we turn to $B$.  First consider $r \gtrsim r_{\rm c}$.  With the boundary condition (\ref{eq:blueshift}) and 
the scalar field solution (\ref{eq:awayfromorigin}),  Eq.~(\ref{eq:Beq})
is solved by
\begin{equation}
\label{eq:Bsol0}
B = -\frac{\kappa v}{2} + O(v^0).
\end{equation}
Next consider $r \lesssim r_{\rm c}$. With the scalar field solution (\ref{eq:nearoriginscalar}),
Eq.~(\ref{eq:Beq}) is solved by
\begin{equation}
\label{eq:Bsol}
B = \frac{\alpha v}{2} \log r + {\rm const.},
\end{equation}
where
\begin{equation}
\alpha > 0.
\end{equation} 
As already noted above, Eq.~(\ref{eq:Beq}) implies $B$ can only decrease as $r$ decreases.  
This means the constant of integration appearing in (\ref{eq:Bsol}) must be $\leq - \frac{\kappa v}{2}$. Therefore, at $r \lesssim r_{\rm c}$
\begin{equation}
\label{eq:Bboundrc}
e^{2 B} \lesssim r^{\alpha v} e^{-\kappa v}.
\end{equation}

Eqs.~(\ref{eq:Bsol0}) and (\ref{eq:Bboundrc}) justify neglecting the ${\textstyle \frac{1}{2}} e^{2 B} \left (1 - {\textstyle \frac{Q^2}{r^2}} \right )$ term in Eq.~(\ref{eq:Veq}).  At late enough times this term is 
order $e^{-\kappa v}$ everywhere, including near $r = 0$.

\subsection{Singularities}

The metric function $B$ is singular at $v = \infty$ and at $r = 0$.  To see that these singularities are physical, consider the Kretschmann scalar. Using the exact equations of motion (\ref{eq:einsteinscalar}) to eliminate derivatives wherever possible, the Kretschmann scalar reduces to
\begin{align}
\nonumber
K &\ =\ e^{-4 B} \bigg \{ 512 \pi^2 (\Psi')^2 (d_+ \Psi)^2  + \frac{12}{r^4} (e^{2B} - 2 V)^2
\\ \nonumber
& \ + \frac{4 Q^2 e^{2B}}{r^8} \left [ e^{2B} (5 Q^2 - 6 r^2) + 12 r^2 V \right ]
\\ \label{eq:Kretch}
& \ + \frac{64 \pi}{r^4} \left [e^{2 B} (r^2 -2 Q^2) - 2 r^2 V \right ] \Psi' d_+ \Psi \bigg \}.
\end{align}

Our late time solutions imply the terms in the braces vanish with
an inverse power of $v$ as $v \to \infty$.  At $r \gtrsim r_{\rm c}$ the dominant term is the first, which vanishes
like $v^{-2 p}$.  This together with  Eq.~(\ref{eq:Bbound}) implies
\begin{equation}
K \sim e^{2 \kappa v} v^{-2 p},
\end{equation}
indicating a null singularity at the CH.
Likewise, Eqs.~(\ref{eq:Bbound}), (\ref{eq:Vsol}) and (\ref{eq:zeta}) imply the mass function
\begin{equation}
\label{eq:massfcn}
m \equiv \textstyle \frac{r}{2} \left [ 1 + \frac{Q^2}{r} - 2 V e^{-2 B} \right ],
\end{equation}
blows up like
\begin{equation}
m \sim e^{\kappa v} v^{-2p},
\end{equation}
which is consistent with well known results from mass inflation \cite{Poisson:1989zz,PhysRevD.41.1796}.  

At $r \lesssim r_{\rm c}$, the scaling relation 
(\ref{eq:Bboundrc}) implies 
\begin{align}
&K \sim r^{-2 \alpha v} e^{2 \kappa v},&&
m \sim r^{-\alpha v} e^{\kappa v}&,
\end{align}
indicating a singularity at $r = 0$.  The strength of the singularity grows 
due to the growing cloud of scalar radiation near $r = 0$ sourced by the Price Law tails.

The above behavior of the Kretschmann scalar is identical to that observed in 
Refs.~\cite{Chesler:2019tco,Chesler:2019pss}.

\subsection{Causal Structure of the spacetime}
\label{sec:causalstruct}

As mentioned above, the outgoing geodesic solution (\ref{eq:outgoingsol})
dictates that geodesics with $r > r_{\rm c}$ terminate at the CH at a finite 
value of $r$ whereas those with $r < r_{\rm c}$ plunge into $r = 0$ in a finite time.
The latter observation implies that the singularity at $r = 0$ must be spacelike.

\begin{figure*}[ht!]
	\includegraphics[trim= 0 0 0 0 ,clip,scale=0.4]{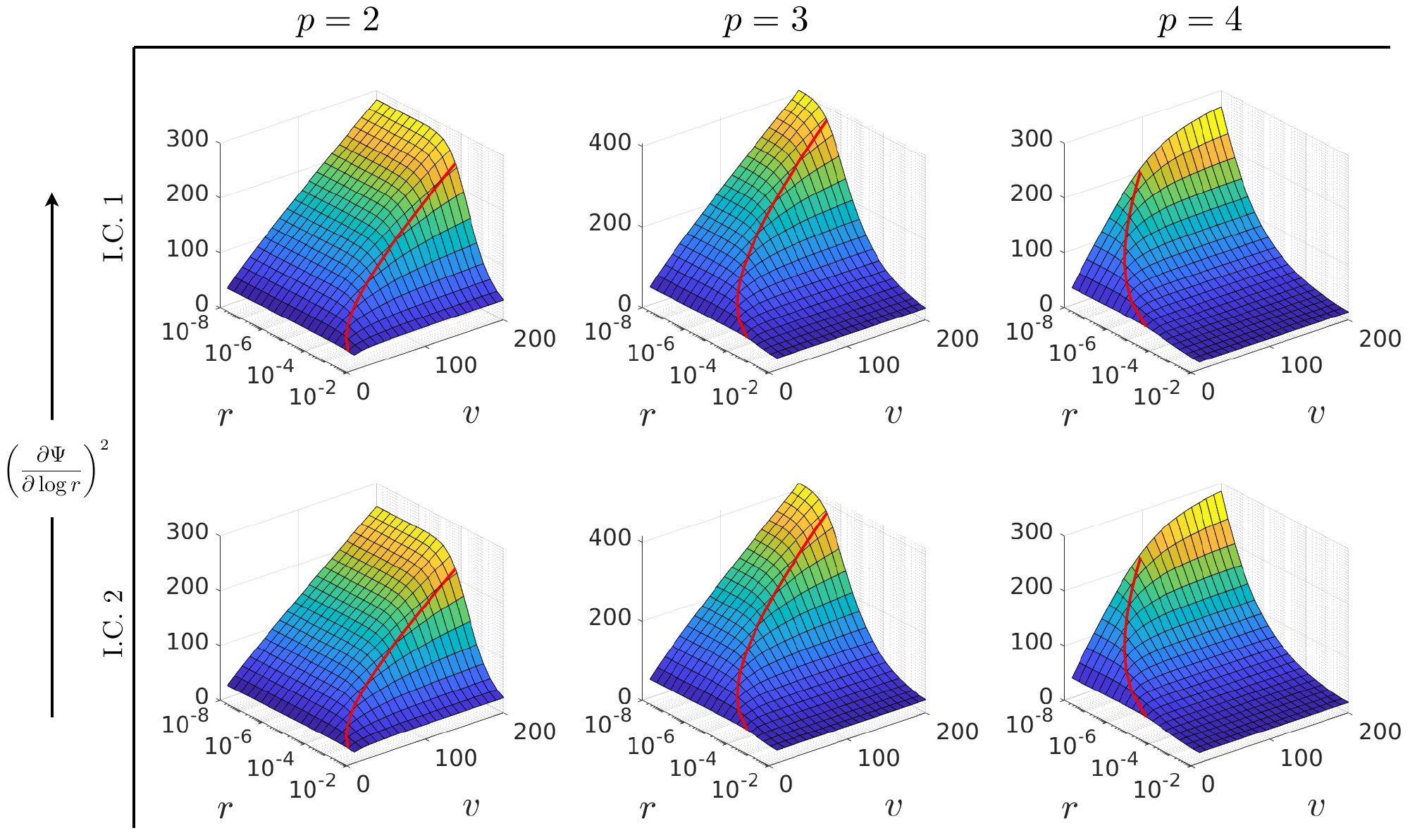}
	\caption{
		$(\partial \Psi/\partial \log r)^2$ for both sets of initial conditions with $p = 2,3$ and $4$ .  The red line superimposed on the plots is $r = r_{\rm c} \sim v^{1/2 - p}$.
		Note $r_c$ decreases more rapidly for larger $p$.
		The numerics are consistent with $\Psi \sim \sqrt{v} \log r$ at $r \lesssim r_{\rm c}$.
	}
	\label{fig:Psi_z}
\end{figure*}

Let us now consider time-like curves.  Demanding  the four velocity has
unit norm means
\begin{equation}
\label{eq:radialvelocity}
\frac{dr}{dv} = - \frac{e^{-2 B}}{(dv/d\tau)^2} + V.
\end{equation}
Here $\tau$ is the proper time of the curve, meaning $\frac{dv}{d\tau}$ is the temporal component of 
the four velocity.  
Just like the null curves discussed above, all time-like curves 
terminate at either $r = 0$ or at a finite value of $r$ at the CH.  
Consider first infalling curves with $\frac{dv}{d\tau} \sim 1$.  Since $e^{-2B} \gtrsim e^{\kappa v}$, these curves have $\frac{dr}{dv} \gtrsim e^{\kappa v}$, and therefore 
terminate at $r = 0$ within proper time $\Delta \tau \lesssim e^{-\kappa v}$.
This is consistent with the Marolf-Ori shock phenomenon \cite{Marolf:2011dj,Eilon:2016osg}.
As argued in Refs.~\cite{Marolf:2011dj,Eilon:2016osg}, upon crossing 
$r = r_-$ infalling observers experience tidal forces of order $e^{2 \kappa v}$ and 
receive an exponentially large kick inwards with $\frac{dr}{d\tau} \sim e^{\kappa v}$.

Time-like curves which terminate at the
CH maximize the proper time.  These curves must have $\frac{dr}{dv} \to 0$ as $v \to \infty$.
Using $e^{-2B} \gtrsim e^{\kappa v}$ and $V \sim -v^{-2p}/r$, Eq.~(\ref{eq:radialvelocity}) implies this condition is satisfied provided
\begin{equation}
\label{eq:dvdtau}
\frac{dv}{d\tau} \gtrsim e^{\kappa v/2}.
\end{equation}
It follows that time-like curves terminate at the CH within proper time
\begin{equation}
\label{eq:Dtau}
\Delta \tau \lesssim e^{-\kappa v/2}.
\end{equation}

Eqs.~(\ref{eq:dvdtau}) and (\ref{eq:Dtau}) merely reflect the exponential blueshift near the CH.  Clocks belonging to observers attempting to cross the CH run exponentially faster than those of the outside universe.
Since
the cutoff $r_{\rm max}$ is arbitrary and the blueshift kicks in at $r < r_-$,
we conclude that all time-like curves inside $r_-$ terminate at a singularity within
proper time (\ref{eq:Dtau}).  Therefore, for time-like observers the classical geometry effectively ends at $r_-$.  Identical conclusions can be reached for the one-sided black holes studied in Refs.~\cite{Chesler:2019tco,Chesler:2019pss}.  Indeed, this conclusion 
should apply to any scenario with mass inflation.

\section{Numerical simulations}
\label{sec:numerics}

\subsection{Setup}

We numerically solve the Einstein-scalar system (\ref{eq:einsteinscalar}) subject to the 
boundary conditions (\ref{eq:blueshift}) and (\ref{eq:Price}).  Based 
on the fact both $\Psi$ and $B$ diverge near $r = 0$ like $\log r$, we employ
\begin{equation}
z \equiv \log r,
\end{equation}
as a radial coordinate.  Likewise, since Eq.~(\ref{eq:Vsol}) implies $V$ diverges like $1/r$, we choose to work with the rescaled variable 
\begin{equation}
\mathcal V \equiv r V.
\end{equation}

Our discretization scheme is discussed at length in Ref.~\cite{Chesler:2013lia}.
We employ pseudospectral methods with domain decomposition with 20 equally spaced domains in $z$.
In each domain we expand the $z$ dependence in terms of the first 8 Chebyshev polynomials.  Derivatives w.r.t. $z$ are defined by differentiating the Chebyshev polynomials.

We have found that when integrating very close to $r = 0$, the equations of motion become very stiff, at least initially.  In fact this initial stiffness is what limits our ability to 
to integrate closer to $r = 0$.  To combat this, we evolve forward in $v$ using Matlab's stiff ODE solver, \textit{ode15s}. 

We fix mass  $M = 1$ and charge $Q = 0.8$,
which via Eqs.~(\ref{eq:rm}) and (\ref{eq:kappa}) yields $r_- = 0.4$ and $\kappa = 3.75$.
We choose Price Law amplitude $A = 0.1$ and powers
$p = 2,3$ and $4$. 
Our radial computational domain is $r \in (r_{\rm min},r_{\rm max})$ with
\begin{align}
&r_{\rm min} = 10^{-8}, & r_{\rm max} = 0.1.&
\end{align}

We begin time evolution at time $v = 2$.
For initial $V|_{r = r_{\rm max}}$ we choose 
\begin{equation}
V|_{r = r_{\rm max}} = -0.01.
\end{equation}
For convenience we focus on initial scalar fields which vanish rapidly near $r_{\rm max}$.
We employ two different initial scalar field profiles,
\begin{align}
&\Psi = \textstyle 10 \left (1 - \frac{r^2}{r_{\rm max}^2} \right)^4,&
\Psi = 10 \textstyle \cos^4 \frac{\pi r}{2 r_{\rm max}}.&
\end{align}
We refer to these two sets of initial conditions as initial condition 1 (I.C.~1)
and initial condition 2 (I.C.~2), respectively. 

Both scalar initial conditions yield large deviations from the RN geometry, with $V < 0$ throughout the entire computational domain.  Note that since $V < 0$, no boundary condition on the scalar field is needed 
at $r = r_{\rm min}$: all excitations propagate towards $r = 0$.  Instead, one must specify the amplitude of outgoing waves at $r = r_{\rm max}$, which also propagate inwards.
The outgoing geodesic equation (\ref{eq:outgoing}) means $dr/dv \sim - v^{-2p}$. Because of this, 
outgoing waves at $r_{\rm max}$ essentially just stay at $r = r_{\rm max}$ and do not affect evolution away from $r_{\rm max}$. For simplicity we set $\Psi'|_{r = r_{\rm max}} = 0$.

\begin{figure}
	\includegraphics[trim= 0 0 0 0 ,clip,scale=0.27]{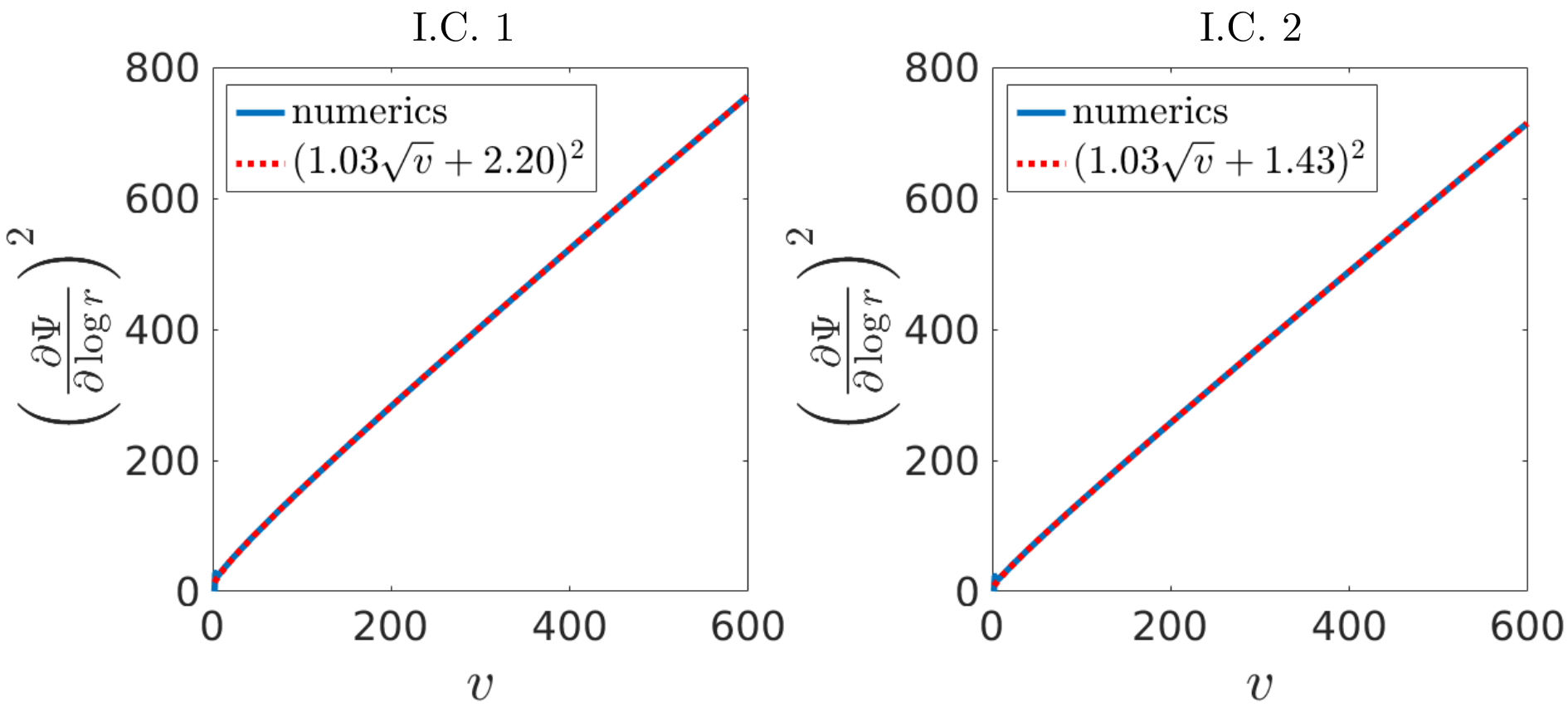}
	\caption{
		$(\partial \Psi/\partial \log r)^2$ at $r = 10^{-8}$ for both sets of initial conditions with $p = 2$. Also included in the plots are fits to $(a \sqrt{v} + b)^2$.  The fit 
		parameter $a$ is the same for both sets of initial conditions whereas $b$ varies by order 50\%.  
	}
	\label{fig:Psi_zFits}
\end{figure}

To test the convergence of our code we monitor violations of the radial constraint 
equation (\ref{eq:constraint}).  In terms of the rescaled variable $\mathcal V$ and the 
radial coordinate $z$, this equation reads 
\begin{equation}
\label{eq:Erreq}
{\rm Err} = 0,
\end{equation}
with
\begin{align}
\nonumber
{\rm Err} \equiv \ & \partial_v \mathcal V - 2  \mathcal V \partial_v B + 4 \pi e^{2 z} (\partial_v \Psi + e^{-2 z} \mathcal V \partial_z \Psi)^2 
\\ \label{eq:constraint2}
&\ - 4 \pi e^{-2 z} \mathcal V^2 (\partial_z \Psi)^2.
\end{align}
Eq.~(\ref{eq:Erreq}) is  enforced exactly at $r = r_{\rm max}$, where it is used
to evolve the boundary value $\mathcal V|_{r = r_{\rm max}}$.  In the continuum 
limit, the remaining equations of motion,  (\ref{eq:Beq}), (\ref{eq:Veq}) and (\ref{eq:scalar2}), dictate ${\rm Err} = 0$
throughout the computational domain.  
In Fig.~\ref{fig:Convergence} we plot 
$|{\rm Err}|$ for I.C.~1 (top row) and I.C.~2 (bottom row)
with $p = 2,3$ and $4$ (left, middle and right columns).  In all simulations we see that 
$|{\rm Err}|$ is largest at very early times.  This is due to the aforementioned initial stiffness of the equations of motion.  Nevertheless, for all simulations we have $|\rm Err| < 2 \times 10^{-5}$ when $v \geq 10$ and $|\rm Err| <  10^{-8}$ when $v \geq 25$.
This indicates our discretized equations of motion well-approximate the continuum limit.

\begin{figure*}[h]
	\includegraphics[trim= 0 0 0 0 ,clip,scale=0.4]{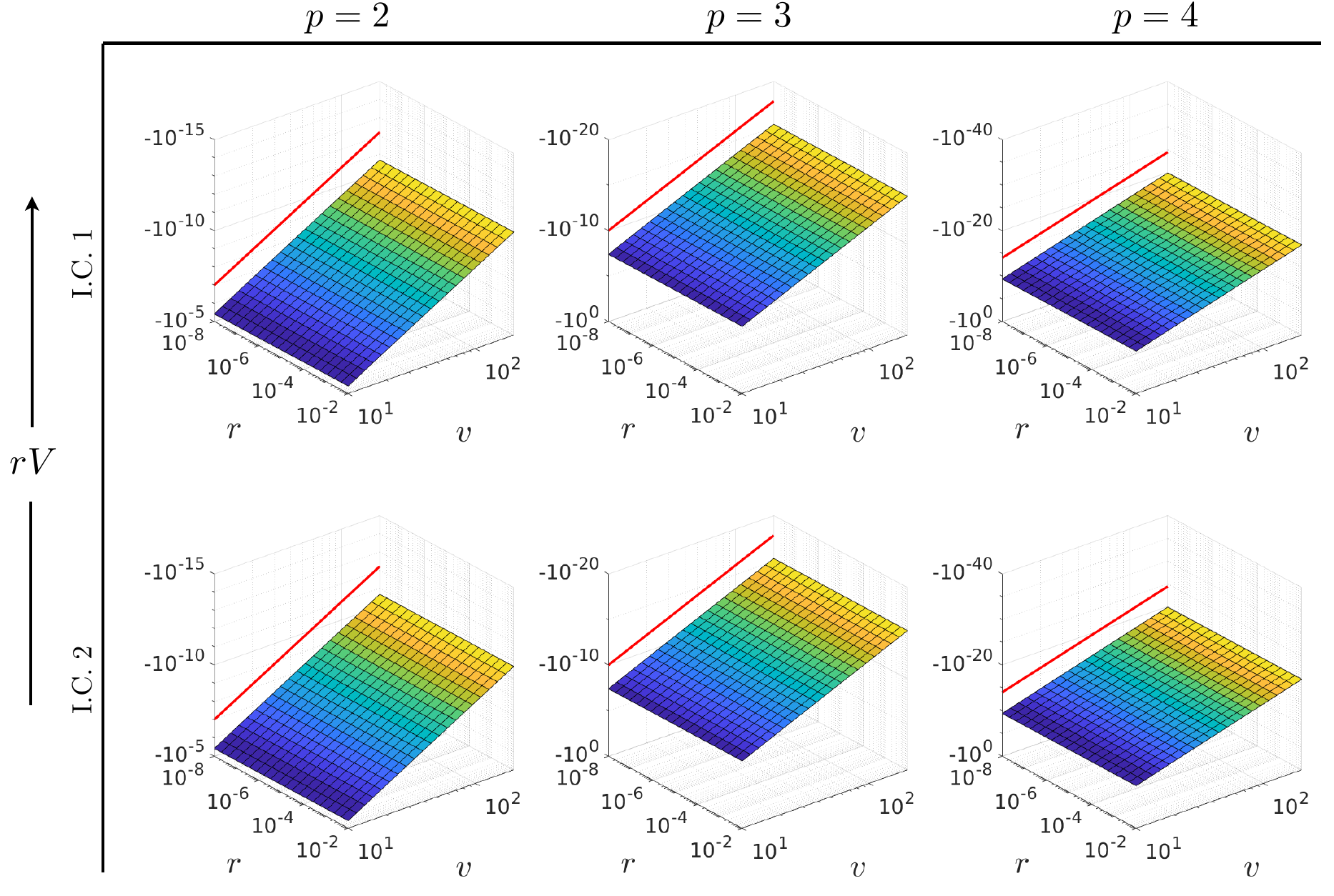}
	\caption{
		$r V$ for both sets of initial conditions for $p = 2,3$ and $4$.  The red line in each plot is $-v^{-2p}$.  The numerical results are consistent with $V \sim -v^{-2p}/r.$
	}
	\label{fig:V}
\end{figure*}

\begin{figure*}[h]
	\includegraphics[trim= 0 0 0 0 ,clip,scale=0.4]{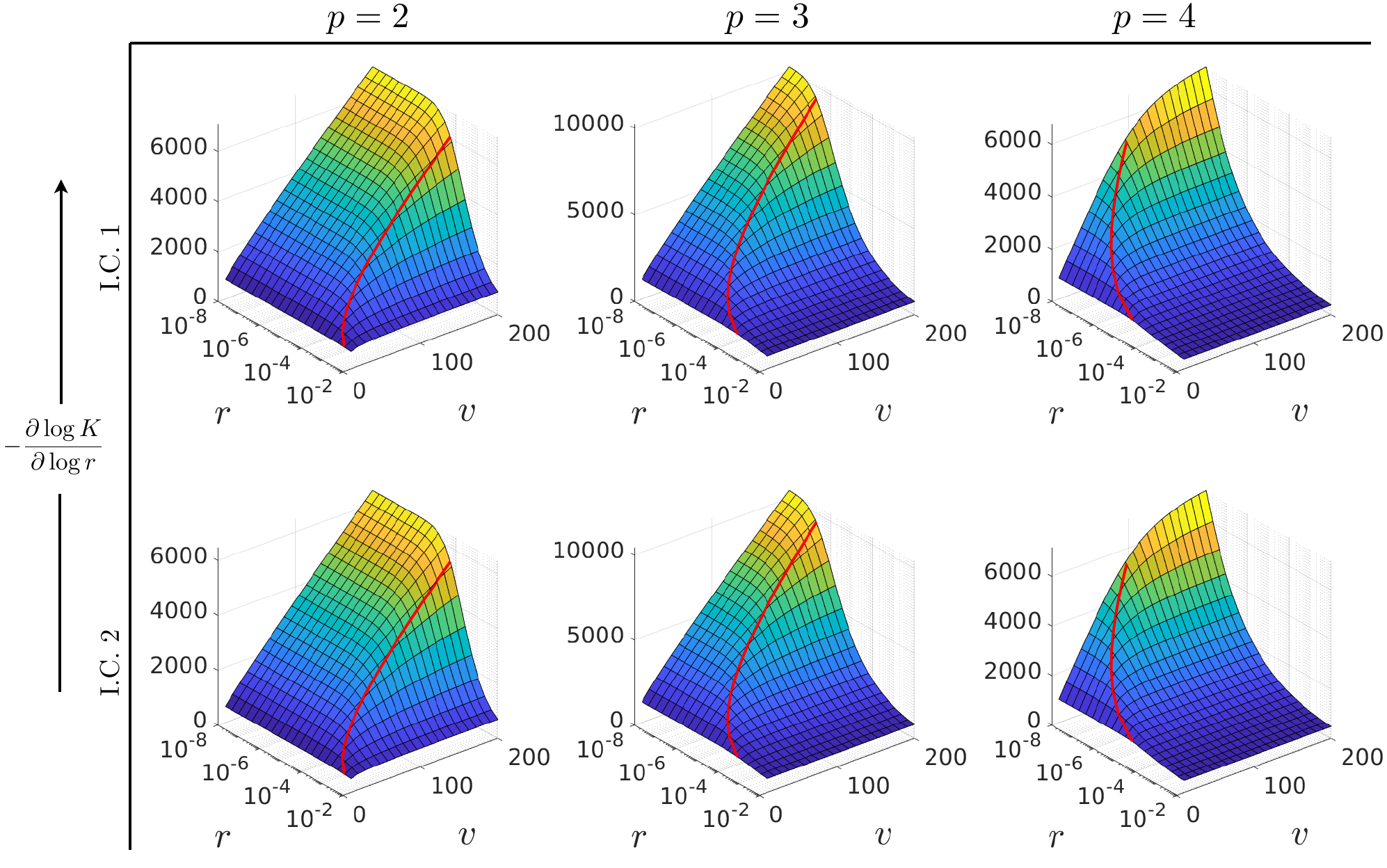}
	\caption{
		Logarithmic derivative of the log of the Kretschmann scalar, $-\partial \log K/\log r$, 
		for both sets of initial conditions with $p = 2,3$ and $4$.  The red line superimposed on the plots is $r = r_{\rm c}  \sim v^{1/2 - p}$.  At $r \lesssim r_{\rm c}$ the numerics are consistent with $K \sim r^{-2 \alpha v}$ for some constant $\alpha > 0$.
	}
	\label{fig:K}
\end{figure*}

\subsection{Results}

\begin{figure*}[ht!]
	\includegraphics[trim= 0 0 0 0 ,clip,scale=0.4]{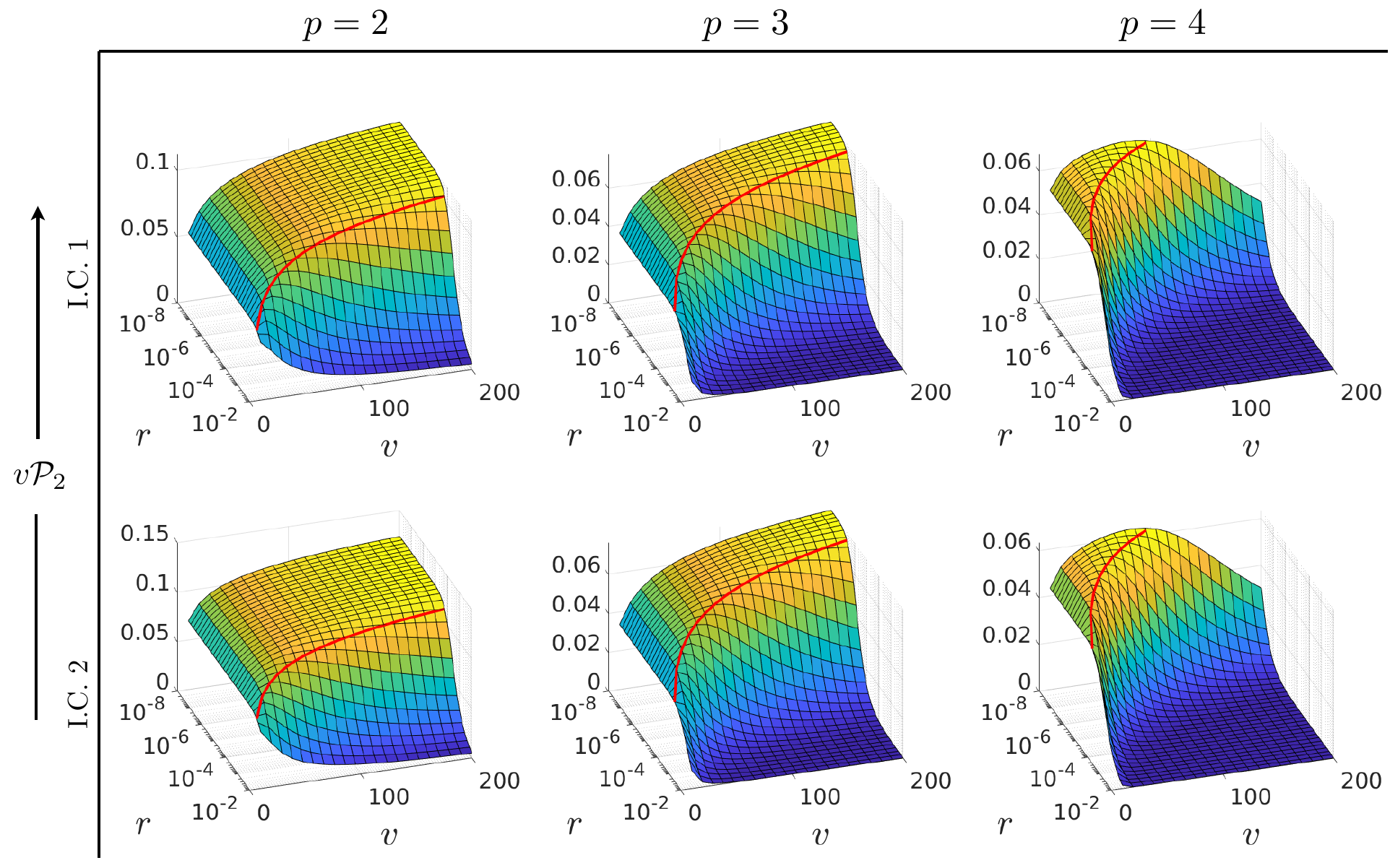}
	\caption{
		$v \mathcal P_2$ for $p = 2,3,4$ and for both sets of initial conditions.  The red line superimposed on the plots is $r = r_{\rm c}$.  At $r \leq r_{\rm c}$ we 
		see $v \mathcal P_2$ plateauing.  The numerics are consistent with $\mathcal P_2 \sim 1/v$ at $r \lesssim r_{\rm c}$.
	}
	\label{fig:P2}
\end{figure*}

In Fig.~\ref{fig:Psi_z} we plot $\left ({\partial \Psi}/ {\partial \log r} \right )^2$ for 
I.C.~1 (top row) and I.C.~2 (bottom row) 
with $p = 2,3,4$ (left, middle, right columns). Superimposed on each plot is the critical radius
$r = r_c \sim v^{1/2 - p}$ (red line).  In all plots we see that at $r \lesssim r_{\rm c}$, 
$\left ({\partial \Psi}/ {\partial \log r} \right )^2 \sim v$.
This behavior is consistent with our late time analysis
in Sec.~\ref{sec:latetimeapprox}, where is was found that $\Psi \sim \sqrt{v} \log r$ when $r \lesssim r_{\rm c}$.  Note this behavior is most pronounced for smaller $p$.  Why? For larger $p$ the curve $r = r_{\rm c}$ decreases more rapidly as $v$ increases.  Indeed, for $p = 4$, $r = r_{\rm c}$ exits our computational domain around time $v = 100$.

As discussed in Sec.~\ref{sec:latetimeapprox}, the $\sqrt{v}$ growth of the scalar field
near $r = 0$ is driven by Price Law influxes.  Recall that in our simulations we chose the Price Law boundary condition (\ref{eq:Price}) 
to be the same for both I.C.~1 and I.C.~2.  This is why the plots of 
$\left ({\partial \Psi}/ {\partial \log r} \right )^2$ in Fig.~\ref{fig:Psi_z} look identical for I.C.~1
and I.C.~2.  Nevertheless, subleading static components of $\Psi$ should be sensitive initial conditions.  Near $r = 0$ and at late times $v$, it is reasonable to expect 
$\Psi \sim (\sqrt{v} + k) \log r$,
where the constant $k$ depends on the initial scalar field profile.
In Fig.~\ref{fig:Psi_zFits} we plot $\left ({\partial \Psi}/ {\partial \log r} \right )^2$ at $r = 10^{-8}$ for  I.C.~1 (left) and I.C. 2 (right), both with  $p  = 2$.  
As is evident from the figure, there is a small amount of curvature in $\left ({\partial \Psi}/ {\partial \log r} \right )^2$, indicating a small deviation from the linear growth $\left ({\partial \Psi}/ {\partial \log r} \right )^2 \sim v$.  Also included in the plots are fits to $(a \sqrt{v} + b)^2$ with fit parameters $a$ and $b$.  The fits agree very well
with the numerical data.  As expected, the fit parameter $a$ is identical for I.C.~1 and I.C.~2.  In contrast, 
$b$ varies by order $50\%$ between I.C.~1 and I.C.~2.

We now turn to the metric functions $V$ and $B$.  Note that by Eq.~(\ref{eq:Beq}), $B' = \frac{2 \pi}{r} \left ({\partial \Psi}/ {\partial \log r} \right )^2$.  Hence Fig.~\ref{fig:Psi_z} implies $B' \sim \frac{v}{r}$ when $r \lesssim r_{\rm c}$.  This is consistent with Eq.~(\ref{eq:Bsol}) in our late time analysis.  In Fig.~\ref{fig:V} we plot $r V$ for 
I.C.~1 (top row) and I.C.~2 (bottom row) 
with $p = 2,3,4$ (left, middle, right columns).  The red line in each plot shows $-v^{-2 p}$.
Eq.~(\ref{eq:Vsol}) in our late-time analysis predicts $V \sim - v^{-2 p}/r$.  As is clear from
Fig.~\ref{fig:V}, all of our simulations are consistent with this result.  The fact that $V< 0 $ 
means the singularity at $r = 0$ must be spacelike.

Finally, in Fig.~\ref{fig:K} we plot  $-\frac{\partial \log K}{\partial \log r}$ for 
I.C.~1 (top row) and I.C.~2 (bottom row) with $p = 2,3,4$ (left, middle and right columns). 
Superimposed on each plot is the critical radius $r = r_{\rm c}$ (red line).  In all plots we see that at $r \lesssim r_{\rm c}$, 
$- \frac{\partial \log K}{\partial \log r}  \sim v$.  This behavior is consistent with our late time analysis in Sec.~\ref{sec:latetimeapprox}, where is was found that $K \sim r^{-2 \alpha v} e^{2 \kappa v}$ when $r \lesssim r_{\rm c}$.  Fig.~\ref{fig:K} clearly demonstrates $\alpha > 0$, meaning the strength of the singularity at $r = 0$ grows with time $v$.


\section{Discussion}
\label{eq:discussion}

Our numerical simulations are completely consistent with 
our late time approximation scheme.  Our analysis demonstrates that 
the late time behavior of the spacelike singularity at $r = 0$ is determined by Price Law influxes,
with the curvature growing according to (\ref{eq:K}). Moreover, 
all time-like curves inside $r_-$ terminate at a singularity at $r = 0$ or $v = \infty$ in an exponentially short proper time.  This means that for time-like observers, the classical geometry effectively ends at $r = r_-$, with a sub-Plackian volume of spacetime lying beyond $r_-$. 

It is instructive to compare our results to expectations from a BKL analysis.  In vacuum the geometry near a BKL singularity is oscillatory and chaotic  \cite{Belinsky:1970ew}.  However, the presence of scalar field ameliorates the oscillatory structure, resulting in a monotonic singularity \cite{Belinski:1973zz}.  On general grounds it is expected 
that in the frame of an infalling observer asymptotically close to the singularity, the metric should only depend on proper time $\tau$ and take the form of the scalarized Kasner metric
\begin{equation}
\label{eq:kasner}
ds^2 = -d \tau^2 + \tau^{2 p_1} dx^2 + \tau^{2 p_2} dy^2 +  \tau^{2 p_3} dz^2.
\end{equation}
With a scalar field the Kasner exponents $p_i$ satisfy $\sum_{i} p_i = 1$, but need not satisfy 
$\sum_i p_i^2 = 1$, as they do for the Kasner geometry.
Note  spherical symmetry dictates two of the exponents are equal, e.g. $p_2 = p_3$.

Using the late time solutions for $V$ and $B$, Eqs.~(\ref{eq:Vsol}) and (\ref{eq:Bsol}), it is straightforward but tedious to find a coordinate transform which takes the Bondi-Sachs metric (\ref{eq:metric}) near $r = 0$ to the scalarized Kasner metric (\ref{eq:kasner}).  Doing so, we find 
\begin{align}
\label{eq:exponents}
&p_1 = \textstyle 1 - \frac{4}{\alpha v} + O(1/v^2), &
p_2 = \textstyle \frac{2}{ \alpha v} + O(1/v^2).&
\end{align}
Thus $p_1 + 2p_2 = 1$ as expected.  The fact that $p_1 \to 1$ indicates 
distances contract in the $x$ direction while remaining constant in the transverse directions as the singularity is approached.

To see that our numerics are consistent with the scalarized Kasner geometry at $r \lesssim r_{\rm c}$, define
\begin{equation}
\rho^{\alpha \beta} \equiv {\textstyle \frac{1}{(\nabla \Psi)^2}}R^{\alpha \mu \beta \nu} \nabla_\mu \Psi \nabla_\nu \Psi,
\end{equation}
and 
\begin{align}
&\mathcal P_1 \equiv \textstyle \frac{R^{\theta \phi}_{ \ \ \ \theta \phi} - R^{\theta}_{\ \theta} + \rho^{\theta}_{\ \theta}}{R^{\theta \phi}_{ \ \ \ \theta \phi} -\rho^{\theta}_{\ \theta}},&
\mathcal P_2 \equiv \textstyle \frac{R^{\theta \phi}_{ \ \ \ \theta \phi}}{R^{\theta \phi}_{ \ \ \ \theta \phi} -\rho^{\theta}_{\ \theta}}. &
\end{align}
Note $\mathcal P_1$ and $\mathcal P_2$
are invariant under coordinate transformations that only mix time and radius. 
For the scalarized Kasner metric (\ref{eq:kasner}),  $\mathcal P_1 = p_1$ and $\mathcal P_2 = p_2$.
We can therefore directly compute $p_1$ and $p_2$ from $\mathcal P_1$ and $\mathcal P_2$ 
in our coordinate system.  Doing so we find $\mathcal P_1 + 2 \mathcal P_2$ virtually indistinguishable from unity in our entire 
computational domain.  Indeed, the equations of motion (\ref{eq:einsteinscalar}) imply 
\begin{equation}
\mathcal P_1 + 2 \mathcal P_2 = 1 + O(e^{2B}),
\end{equation}
meaning up to exponentially small corrections, $\mathcal P_1 + 2 \mathcal P_2 = 1$ everywhere in our computational domain.  

In Fig.~\ref{fig:P2} we plot $v \mathcal P_2$ for 
I.C.~1 (top row) and I.C.~2 (bottom row) 
with  $p = 2,3,4$ (left, middle, right columns). Superimposed on each plot is the critical radius
$r = r_{\rm c}$ (red line).  In all plots we see that at $r \lesssim r_{\rm c}$, $v \mathcal P_2 \approx {\rm const.}$  This is consistent a scalarized Kasner geometry at $r \lesssim r_{\rm c}$,
with Kasner exponent $p_2 \sim 1/v$.

Thus far we have focused solely on the case where the geometry  is regular at $r > 0$.
Nevertheless, it is possible to apply the late time approximation when there is a singularity 
at $r = r_{\rm CH} > 0$.  In fact,  the late time approximation improves because of the singularity.
Specifically,  if the scalar field $\Psi$ is more singular than $(r - r_{\rm CH})^{1/2}$, then 
the Einstein equation (\ref{eq:Beq}) implies $B \to - \infty$ as $r \to r_{\rm CH}$.  This in turn means the approximation used to obtain Eq.~(\ref{eq:Veq2}) --- namely 
neglecting the terms in Eq.~(\ref{eq:Veq}) which are proportional to $e^{2 B}$ --- becomes better and better as $r \to r_{\rm CH}$ since $e^{2 B}$ vanishes there.

Evolution of the scalar field and its singularity are governed by the decoupled linear Klein-Gordon equation (\ref{eq:scalar3}).
Near $r = r_{\rm CH}>0$ the scalar field becomes arbitrarily rapidly varying, meaning terms in Eq.~(\ref{eq:scalar3}) with first order derivatives
can be neglected yielding 
\begin{equation}
(d_+ \Psi)' = 0,
\end{equation}
where as usual $d_+ = \partial_v - \frac{\zeta(v)}{r} \partial_r$ is the directional derivative along outgoing null geodesics.
This equation, which is just the equation of motion of geometric optics, is solved by 
\begin{align}
\Psi(v,r) = f(v) + h\! \left(r^2 {-} r_{\rm CH}(v)^2 \right),
\end{align}
where $f$ and $h$ are arbitrarily functions determined by initial and boundary conditons, and $r_{\rm CH}(v)$ 
is given by the outgoing geodesic equation (\ref{eq:outgoingsol}), meaning 
$r_{\rm CH}(v)$ is an outgoing null surface.
Moreover, since geodesics inside $r_{\rm c}$ terminate at $r = 0$ in a finite time, it follows that 
if $r_{\rm CH} < r_{\rm c}$, then $r_{\rm CH}$ must contract to $r = 0$ in a finite time,
resulting in the formation of a spacelike singularity.
Conversely, if $r_{\rm CH} > r_{\rm c}$, then the left and right branches of the CH intersect at a finite value of $r$.
It would be interesting but challenging to simulate such scenarios numerically.  We leave this for future work.

\begin{acknowledgments}
This work was supported by the Black Hole Initiative at Harvard University, 
which is funded by the John Templeton Foundation and the Gordon and Betty Moore Foundation. 
We thank Amos Ori and David Garfinkle for useful discussions.
\end{acknowledgments}

\bibliographystyle{utphys}
\bibliography{refs}%

\end{document}